\documentclass[conference]{IEEEtran}

\usepackage{verbatim}
\usepackage{times}
\usepackage{balance}
\usepackage[dvipsnames]{xcolor}
\usepackage{amsmath}
\usepackage{balance}
\usepackage{xspace}
\usepackage[hyphens]{url}
\usepackage{array}
\usepackage{multirow}
\usepackage{subfig}
\usepackage{hyperref}
\usepackage{setspace}

\usepackage[pdftex]{graphicx}
\DeclareGraphicsExtensions{.pdf,.png,.jpg}
\graphicspath{{./plots/}{./liar_plots/}}

\newcommand{\one}{({\em i}\/)}
\newcommand{\two}{({\em ii}\/)}
\newcommand{\three}{({\em iii}\/)}

\def\eg{\emph{e.g.,}\xspace}
\def\etc{\emph{etc.}\xspace}
\def\ie{\emph{i.e.,}\xspace}

\def\vs{\emph{vs.}\xspace}

\usepackage{ifthen,amssymb}
\newboolean{showcomments}
\setboolean{showcomments}{true} 
\ifthenelse{\boolean{showcomments}}
{ \newcommand{\mynote}[3]{
    \fbox{\bfseries\sffamily\scriptsize#1}
    {\small$\blacktriangleright$\textsf{\emph{\color{#3}{#2}}}$\blacktriangleleft$}}}
{ \newcommand{\mynote}[3]{}}

\makeatletter
\def\@IEEEpubidpullup{8\baselineskip}
\makeatother

\begin{document}

\IEEEoverridecommandlockouts
\IEEEpubid{
\parbox{\columnwidth}{\vspace{-4\baselineskip}Permission to make digital or hard copies of all or part of this work for personal or classroom use is granted without fee provided that copies are not made or distributed for profit or commercial advantage and that copies bear this notice and the full citation on the first page. Copyrights for components of this work owned by others than ACM must be honored. Abstracting with credit is permitted. To copy otherwise, or republish, to post on servers or to redistribute to lists, requires prior specific permission and/or a fee. Request permissions from \href{mailto:permissions@acm.org}{permissions@acm.org}.\hfill\vspace{-0.8\baselineskip}\\
\begin{spacing}{1.2}
\small\textit{ASONAM '17}, July 31 -August 03, 2017, Sydney, Australia \\
\copyright\space2017 Association for Computing Machinery. \\ACM ISBN 978-1-4503-4993-2/17/07?/\$15.00 \\
\url{http://dx.doi.org/10.1145/3110025.3110075} \end{spacing}
\hfill}
\hspace{0.9\columnsep}\makebox[\columnwidth]{\hfill}}
\IEEEpubidadjcol

\title{Fake it till you make it:\\Fishing for Catfishes}

\author{
\IEEEauthorblockN{
Walid Magdy\IEEEauthorrefmark{1},
Yehia Elkhatib\IEEEauthorrefmark{4},
Gareth Tyson\IEEEauthorrefmark{3},
Sagar Joglekar\IEEEauthorrefmark{2}, and
Nishanth Sastry\IEEEauthorrefmark{2}
}
\IEEEauthorblockA{
\IEEEauthorrefmark{1}School of Informatics, The University of Edinburgh, UK\\
\IEEEauthorrefmark{4}School of Computing and Communications, Lancaster University, UK\\
\IEEEauthorrefmark{3}School of Electronic Engineering and Computer Science, Queen Mary University of London, UK\\
\IEEEauthorrefmark{2}Department of Informatics, King's College London, UK\\
\tt wmagdy@inf.ed.ac.uk,
y.elkhatib@lancaster.ac.uk,
gareth.tyson@eecs.qmul.ac.uk,\\
sagar.joglekar@kcl.ac.uk,
nishanth.sastry@kcl.ac.uk
}}

\maketitle

\begin{abstract}
Many adult content websites incorporate social networking features.
Although these are popular, they raise significant challenges, including the potential for users to ``catfish'', \ie to create fake profiles to deceive other users.
This paper takes an initial step towards automated catfish detection. We explore the characteristics of the different age and gender groups, identifying a number of distinctions. Through this, we train models based on user profiles and comments, via the ground truth of specially verified profiles. When applying our models for age and gender estimation to unverified profiles, 38\% of profiles are classified as lying about their age, and 25\% are predicted to be lying about their gender.
The results suggest that women have a greater propensity to catfish than men. 
Our preliminary work has notable implications on operators of such online social networks, as well as users who may worry about interacting with catfishes.
\end{abstract}


\section{Introduction}

Adult content has been a long standing innovator in technology. Whereas, originally, online adult content was primarily distributed via pay-per-view sites and peer-to-peer networks, we have recently witnessed a radical shift termed \emph{``Porn 2.0''} with the integration of diverse features into popular portals, \eg videos, images, webcams and chat functionality. For example, many now support user-generated content (UGC), as well as video commenting and rating. These services have exploded in popularity, yet research has not kept pace with their advancement. 

One of the most powerful features introduced is that of online social networking (OSNs). Adult OSNs, much like Facebook, allow users to create and interconnect social profiles. Anecdotally, these profiles have led to a  plethora of fake accounts created by users in an attempt to deceive others regarding their true identity --- so called ``catfishing''~\cite{Cooper2000}. In this paper, we restrict ourselves to two forms of identity deception: users lying about either their gender or age (or both). Although we do not take a moral position about catfishing and perhaps in many cases it does not harm users, there is also a significant subset of situations in which such fake accounts could directly damage others (or themselves). For example, catfishes may pretend to be younger than their true age in an attempt to attract partners; similarly, younger users may pretend to be older to avoid legal age filters. Alternatively, men may pretend to be (homosexual) women in an attempt to garner more female friends~\cite{tyson2015people}. Indeed, our exploration of one such Porn 2.0 OSN userbase reveals that certain demographics do excel in terms of popularity metrics (\S\ref{sec:char}), potentially motivating deceit.
Thus, any mechanism to detect such deception would have significant value to both the OSN operators and users concerned about being deceived.

In this paper we ask \emph{to what extent does catfishing occur}, and also \emph{is it possible to automatically detect them?}
As a first step towards answering this, we study the PornHub adult OSN. We have crawled the PornHub website recording all data from 99,727 OSN profiles (\S\ref{sec:data}). PornHub incorporates some of the most sophisticated social networking features seen in the domain, including the ability to form friendships, upload and share content, send messages and post on each others' ``walls''. This makes it ideal as a case-study for analysis.

A particularly novel feature of PornHub is the ability to create ``verified'' profiles that are manually checked by PornHub employees. This gives a unique ground truth of accurate profiles. Using this data, we characterise the activities of these different genders and age groups to observe notable differences (\S\ref{sec:char}). With these analyses, we move on to explore several algorithms to detect users lying about their age and/or gender (\S\ref{sec:predict}). Using well recognised state-of-the-art techniques~\cite{age:nguyen2011author,age:nguyen2014gender}, and with an additional set of features, we build models to predict true user demographics (gender and age). These predictions allow us to identify unverified users whose listed demographics deviate from the prediction (\ie users who are potentially catfishing). Although this approach lacks a ground truth and should therefore be treated with caution, it does provide a first step towards exploring trends.

With this subset of profiles classified as catfishes, we then begin to explore their characteristics (\S\ref{sec:results}). 
We compute popularity metrics (\eg number of friends) for users predicted to be catfishes to find that they tend to gain larger friendship and subscription groups. Although this metric may be skewed by the possibility of catfishes being more proactive in befriending others, we also note that catfishes gain more profile views as well (a metric which is more difficult to game as it depends on others choosing to visit your page). We surmise that the increasing integration of our offline and online personas will result in this becoming even more important (\S\ref{sec:conc}). Of course, although catfish detection may be more explicitly required in adult portals, our methods will also have benefits for various other services, \eg online dating and mentorship.

The contributions and roadmap of this work is as follows:
\begin{itemize}
\item We present the first work to explore the presence of catfishes on adult OSNs (\S\ref{sec:char}), making our dataset available for public use\footnote{\url{http://homepages.inf.ed.ac.uk/wmagdy/resources.htm}} (\S\ref{sec:data}).
\item We build classifiers for age estimation and gender prediction for adult OSN users, and compare various features for achieving the best performance (\S\ref{sec:predict}).
\item We profile the demographics and activities of users classified as catfishes, highlighting how they differ from other users, and benefit from their deception (\S\ref{sec:results}).

\end{itemize}

\section{Related work}
\label{sec:rw}
Pornography is anecdotally the most searched for content on the web. Whereas work has gone into understanding the sociological aspects of sexual activities \cite{carroll2008generation,daneback2012outcomes}, little is known about the online engines that enable its distribution, especially the expanding ``Porn 2.0'' phenomenon. Numerous YouTube-like websites have emerged (\eg PornHub, YouPorn), with in-built social features. Recent work~\cite{Tyson:2016} recorded over 60 billion views on one such Porn 2.0 website, whilst another study found that some adult video sites can even exceed the traffic footprint of traditional video sites~\cite{Fiadino13}.

A novel requirement that stands out in this domain is the need for robust demographics verification. This is particularly driven by the potential incentives for people to deceive others about their true gender and age. There have been a number of recent studies looking at the automated verification of demographics \cite{age:buzzell2005demographic,age:marquardt2014age,age:rangel2013use}.
For instance, \cite{age:nguyen2011author,age:nguyen2013old} utilised textual features including text content, part-of-speech (POS) tags, and discourse styles to train a model using linear regression for age prediction. They tested their approach on three different collections including online forums. They managed to achieve an age prediction with correlation 0.535 and mean-absolute-error (MAE) in age of 6.5 years. In an extension to their work~\cite{age:nguyen2014gender}, they applied their methods on Twitter data, and achieved a MAE of 4.1, which was shown to be better than human estimation of age.
In another study, \cite{age:al2012homophily} utilised profile and network information for age and gender prediction. The results show that network information leads to significant improvement in prediction, and they explain it as a reason of the principle of homophily of Twitter users.
Unlike the above studies, We do not strive to make a contribution to the algorithmic detection of users lying about demographics. Instead, we exploit these existing state-of-the-art methods and apply them to this new and important domain. This is a topic of increasing importance, with countries such as the UK proposing mandatory age-checks for such websites~\cite{UKporn}. 

To the best of our knowledge, there has been no prior work focusing on demographic verification within the adult domain, although various works have focused on related components of Porn 2.0, such as pornographic practices, communities and subcultures~\cite{attwood2010porn}; interest recommendations~\cite{schuhmacher9exploring}; dating services~\cite{jacobs2009,tyson2016first}; user commenting \cite{trestian2013}; content popularity~\cite{Tyson13} and illegal content dissemination~\cite{Hurley13}. The closest to our work is a recent study of the PornHub social network~\cite{tyson2015people}, although it did not touch upon age verification.

A number of studies have also looked at online gender swapping more generally~\cite{bruckman1996gender}. For example,~\cite{martey2014strategic,lou2013} found that gender swapping in online gaming is commonplace. The reasons for such activities have also been explored via surveys~\cite{hussain2008gender}. Reasons given include: \one~curiosity and the desire to experiment; \two~the perception that the opposite gender is treated better; and \three~the belief that playing the opposite gender will allow new forms of behaviour, or gain advantages. This complements our own work, although we do not focus on finding the reasons for users to catfish. 
Other general work on identifying deception in online communication include linking authors across OSNs \cite{almishari2016trilateral}, detecting online personas \cite{6459494}, and identifying divergent political inclinations \cite{rowe2016mining}. 


\section{Methodology \& Data}
\label{sec:data}

\begin{figure*}
	\centering
 	\fbox{\includegraphics[width=0.7\textwidth]{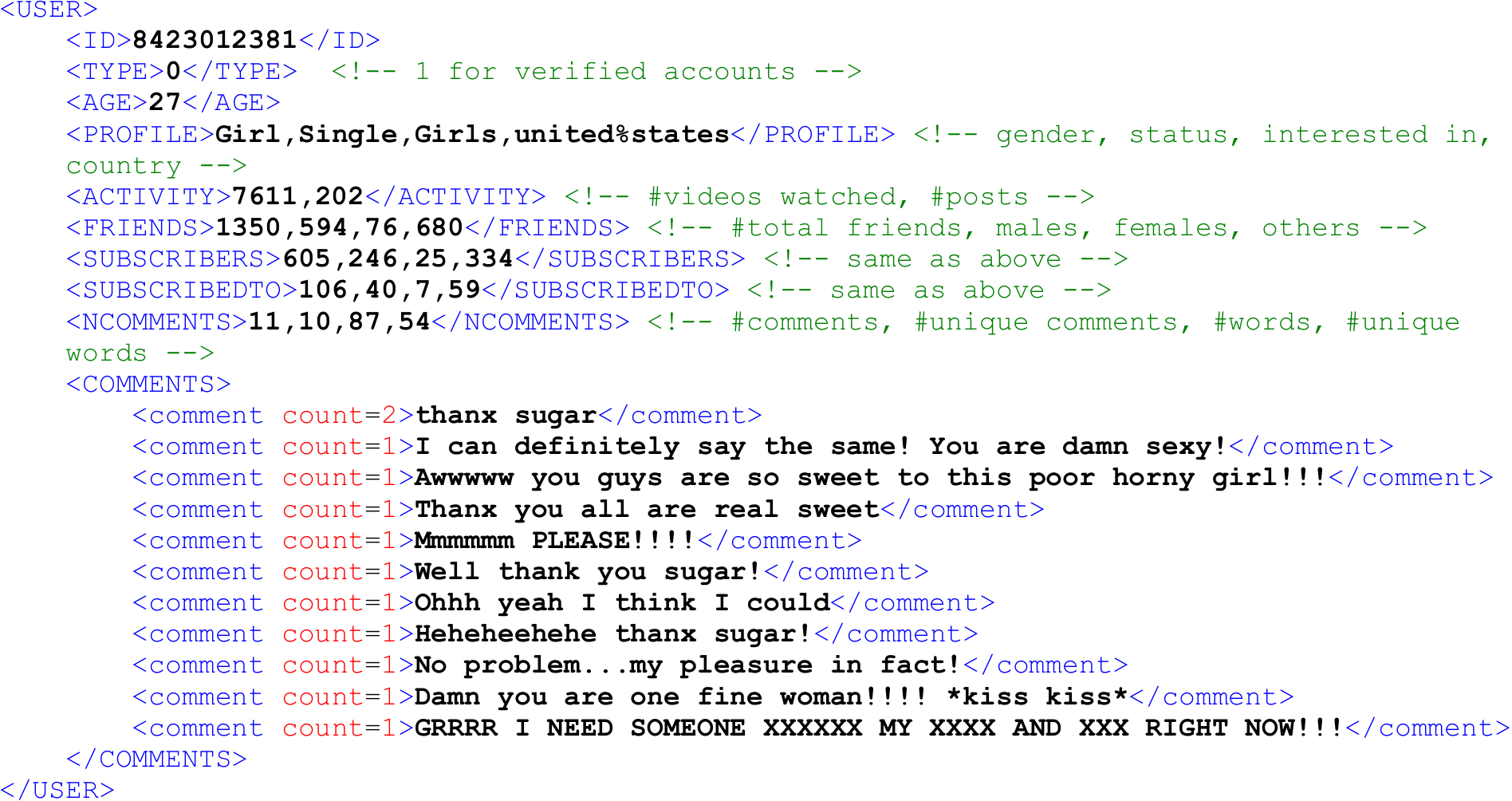}}
    \caption{A sample account in the collected data. Pornographic terms are replaced by X's.}
    \label{fig:data_sample}
\end{figure*}

\subsection{Data Collection}
PornHub is a video sharing website that allows users to upload and view adult content, much like YouTube. It also has a built-in OSN, allowing users to create profiles and form social connections (friendships and subscriptions) with each other. They can exchange private messages, upload/share videos and pictures and post on each others' ``walls''. 

We have scraped the associated OSN profiles attached to PornHub. To initiate the crawl, we used the search facilities on PornHub to retrieve all users from the 60 largest cities within the US, as well as 48 capital cities selected around the world, \eg Beijing, Berlin, London, Tokyo, \etc 
Starting from this seed set, we performed a breadth-first crawl (traversing friendship links). If not already available, we also collected the account information for any users who had left a comment on another user's wall.
In total, we collected 99,727 profiles. 
In each profile, where available, we recorded user age, gender, sexual preference, their ``wall'' of comments, number of profile views, number of videos that they have watched, and their social relationships. Figure~\ref{fig:data_sample} presents a sample account from our data collection showing the information gathered for each account. User IDs in our collection were anonymised to protect users' privacy.

We collect two profile types. The most populated is what we term ``unverified'', which any user can set up without any pre-requisites (\eg verification, payment). These accounts constitute 94.5\% of all users. There is, however, a special form of user account, called ``verified''. These accounts have been manually validated by PornHub; this involves taking a picture of oneself with some message that links the individual to the account (\eg next to their username). Although not foolproof, it does lend a far higher level of user trust in profile details (particularly gender and age) than typical. To increase our number of verified profiles, we specifically searched for all verified profiles in the above cities, and scraped their accounts. Overall, 5.5\% of the accounts crawled were verified (5,484 accounts in total). Despite being a smaller sample size, verified accounts offer a more reliable view of dedicated users, providing a form of (partial) ground truth against which statistics about unverified accounts can be compared. We later exploit these ``ground truth'' verified accounts to train models that can be used to classify the remaining unverified accounts.

\subsection{Limitations \& Ethical Considerations}

Before continuing, it is important to recognise the limitations of our data. First, we acknowledge that the PornHub userbase is not necessarily representative of the wider population. Hence, our results are specific to the PornHub social network.
Second, we later utilise verified profiles to train age and gender prediction models (for use on unverified profiles). An obvious possibility is that the verified profiles and unverified profiles are sufficiently different to make comparability difficult. Similarly, deceptive users may actively modify their behaviour to better reflect their chosen gender and age. Consequently, due to the lack of a comprehensive ground truth, profiles classified as catfishes may be susceptible to significant noise. Differences in population sizes for different demographics (\eg men \vs women) may further undermine the results. It is therefore important to emphasise that results are preliminary and require validation. \emph{We temper all our analysis with this consideration}. Further exploration is a major theme within our future work.

Finally, it is important to note that we are collecting personal information about a large number of users. Hence, we took a number of steps to address ethical concerns. We avoided analysing personally identifiable information, \eg names. We also ensured that our methodology did not involve any communications with the accounts under study. Critically, PornHub allows users to set their profiles to either public or private. Our data is exclusively made-up of users who have set their profiles to public (\ie anyone can view them). We did not make any attempt to access private profiles, which would have involved befriending others.

\section{Characterising PornHub Demographics}
\label{sec:char}
Before exploring the presence of catfishes, we briefly characterise the demographic make-up of PornHub accounts. We include both verified and unverified accounts in this analysis, but we ignore accounts labelled as ``couple'', ``company'', ``transgender'', or ``not specified''' (these collectively make just 4.4\% of all accounts) to focus on catfishing in the context of male and female accounts. 
Note that we interchangeably use the terms `girl' / `guy' and female / male as these are the terms used on the portal.

\begin{figure}
	\centering
 	\includegraphics[width=0.98\columnwidth]{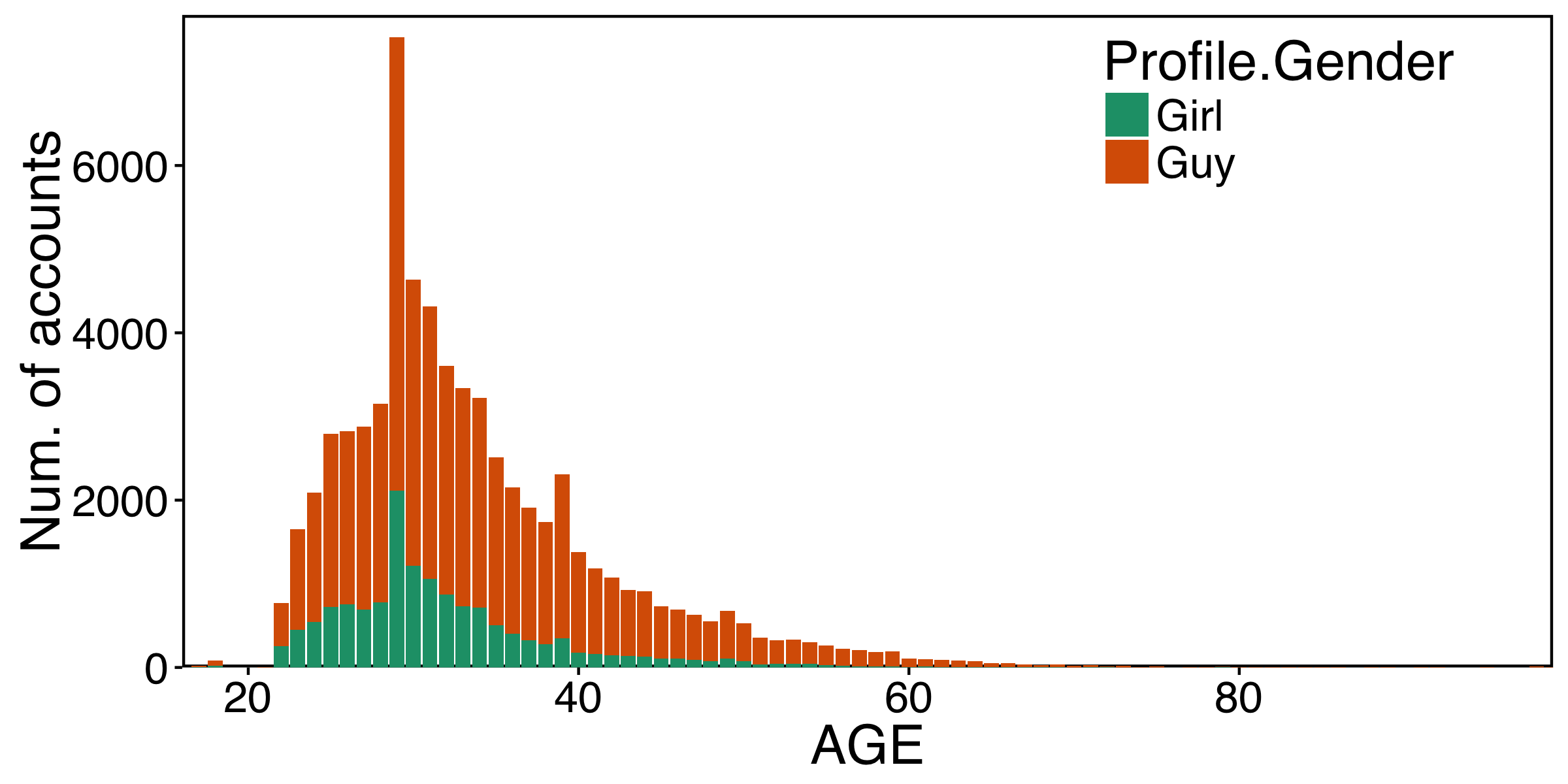}
    \caption{Distribution of male and female profiles across age ranges}
        \label{fig:age_hist}
\end{figure}

Figure~\ref{fig:age_hist} presents a histogram revealing the number of accounts of each age. 64\% of accounts fall into the 18--30 bracket, confirming a young demographic. Perhaps more noteworthy is the distinction between male and female accounts, with female users marginally younger (average of 28 \vs 31). This is driven by a larger population of older male profiles.  This could perhaps be explained by that fact that younger women have shown greater affinity to pornographic content (compared to older), whilst men continue to maintain an interest into older years~\cite{ferree2003women}. 

\begin{figure}
	\centering
 	\includegraphics[width=0.98\columnwidth]{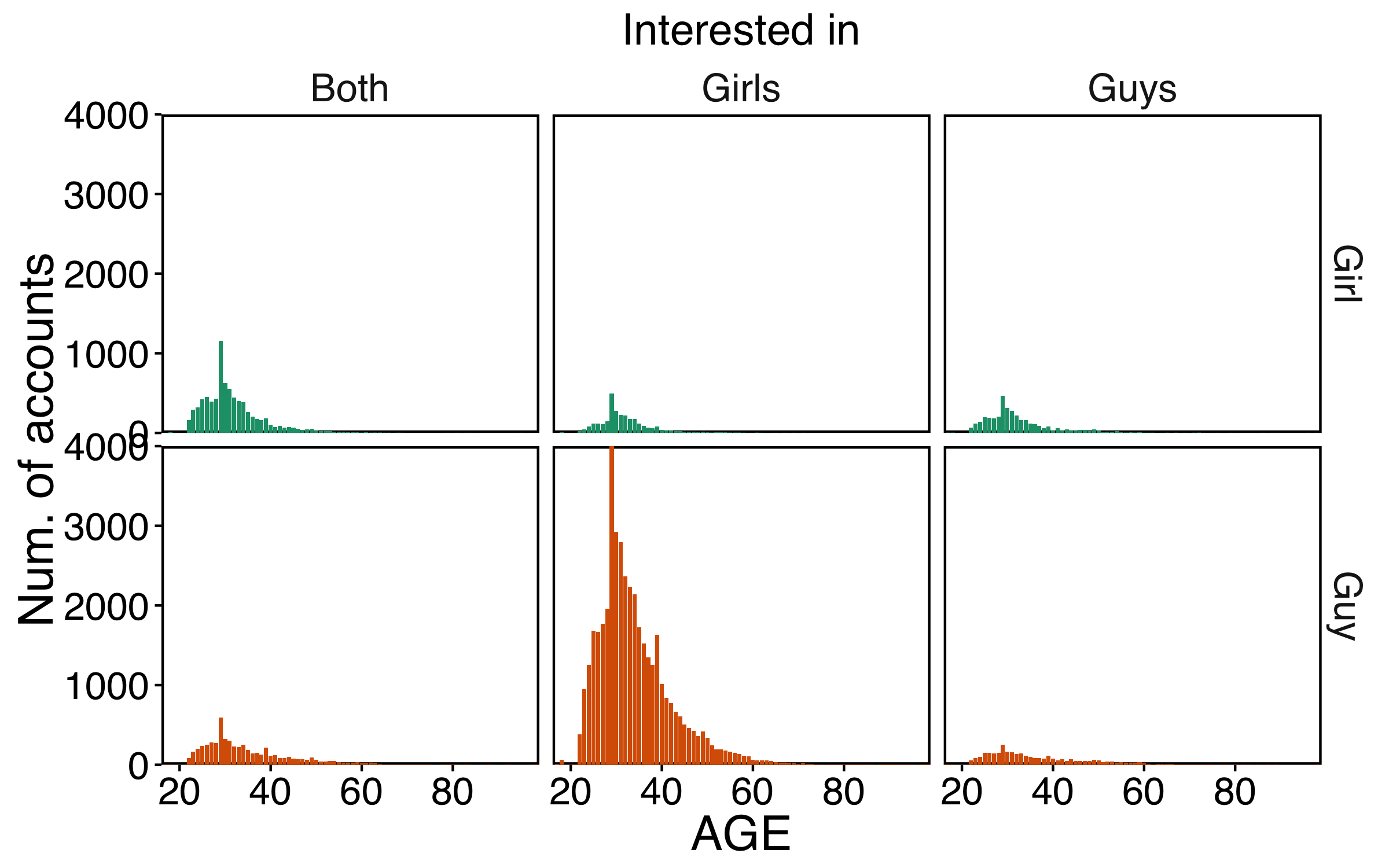}
    \caption{Distribution of male and female users interested in men, women and both across age ranges}
        \label{fig:histo_age_interest}
\end{figure}

As well as being able to stipulate their own gender, profiles can also list which other gender(s) they are interested in. Figure~\ref{fig:histo_age_interest} presents the number of accounts that stipulate an interest in each gender (male, female, both). We find that 68\% of accounts are interested in women (bottom middle plot) of which 90\% are heterosexual males, yet only 4.5\% of accounts (top right plot) are females interested in males who could satisfy this need. 
In fact, female accounts are \emph{more} likely to stipulate an interest in both, rather than men alone.
In other words, PornHub suffers from a significant supply-and-demand mismatch: Female profiles are in high demand, yet only a small proportion of profiles can satisfy this demand.

\begin{figure}
	\centering
 	\includegraphics[width=0.98\columnwidth]{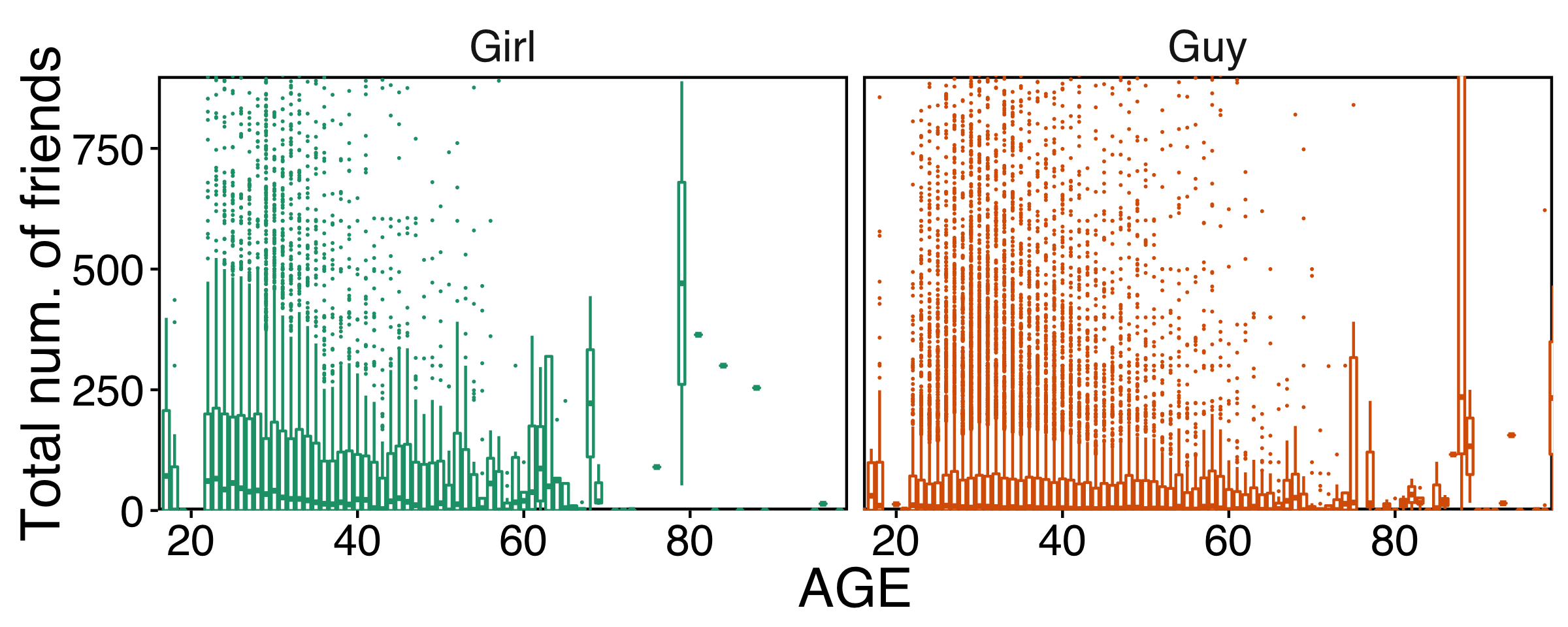}
    \caption{Number of friends per profile \vs reported age}
    \label{fig:box_age_friends_gender}
\end{figure}

To explore this further, we look at the number of friends garnered per user for both female and male accounts (Figure~\ref{fig:box_age_friends_gender}). Profiles labelled as female consistently obtain larger social groups than male ones: an average of 396 \vs 185. 
A downward trend can also be observed across the age ranges with younger users gaining more friends, which is again more pronounced for female accounts. For example, the average 21 year old female accounts gets 433 friends compared to just 176 for a 50 year old female.

One explanation for the above could be that younger users are more proactive in befriending others. To explore this, we inspect the number of subscribers for each account type (these are equivalent to followers on Twitter, \ie directed relationships). Again, we see a strong bias towards female profiles who, on average, have 309 subscribers \vs just 97 for men. Similarly, the number of friends, on average, for female accounts is 403 \vs 188 for men. Clearly, the subscriber numbers are primarily driven by the behaviour of \emph{other} users (as a user cannot subscribe to themselves). Hence, this adds evidence against the claim that young female profiles may get large social groups simply through proactive befriending other. In fact, the average female account is subscribed to just 106 channels compared to 142 for male accounts.

In summary, the above confirms a significant difference between male and female profiles in terms of social popularity (friendship groups). Our aim is not to speculate on \emph{why} this is the case. However, it is clear that these demographic pressures might create incentives for people to lie about profile details. Due to this, we spend the rest of the paper exploring the propensity for users to lie on this social network.






\section{Demographic Prediction}
\label{sec:predict}

To identify deceptive users (catfishes), it is first necessary to extract and predict profile features that indicate true demographics. To do this, we utilise PornHub's \emph{verified profiles}; these are profiles that have been independently verified by PornHub.
Although an assumption, these are significantly more trustworthy than non-verified accounts, since they must pass a process of validation by the website.

We use the verified profile data as a training set for building models for age estimation and gender prediction. Models are then used to estimate the demographics of the unverified accounts and compare them to the claimed ones. It is worth noting that verified users may exhibit different characteristics (\eg activity level, sexual preferences, education level) to their unverified equivalents, thereby making their use within the classifier less accurate. To mitigate this, we only identify users as catfishes when they exhibit a significant disparity to their predicted demographics. Regardless, we emphasise that the \emph{lack of a ground truth means that our results should only be treated as a first-stage analysis.}

\def\arraystretch{1.25}

\begin{table*}
\begin{center}									
{	
\begin{tabular}{l|ccc|ccc|cc}
Features & \multicolumn{3}{c}{Female}	& \multicolumn{3}{c}{Male} & Macro F1 & Accuracy \\\hline
& R & P & F1 & R & P & F1 & & \\\hline
Content & 0.542 & 0.885 & 0.672 & 0.974 & 0.854 & 0.910 & 0.791 & 0.859 \\
Network & 0.702 & 0.847 & 0.768 & 0.954 & 0.898 & 0.925 & 0.846 & 0.887 \\
All & \textbf{0.769} & \textbf{0.920} & \textbf{0.838} & \textbf{0.976} & \textbf{0.921} & \textbf{0.947} & \textbf{0.893} & \textbf{0.920} \\\hline
\end{tabular}
\caption{Gender prediction performance}
\label{tab:GenderResults}}	
\end{center}
\end{table*}

\subsection{Feature Extraction}
First, a set of features is extracted for training a regression model for age estimation and a classifier for gender prediction. Many features are inspired by the state-of-the-art~\cite{age:nguyen2011author,age:nguyen2014gender,age:al2012homophily}; however, we also introduce a set of novel features that are specific to our data. 
The set of feature groups extracted from the data set are as follows:
\subsubsection{Content features}
The comments collected for each user's profile are used for generating a set of features. Some of the content features are inspired by~\cite{age:nguyen2011author}, while we introduce an additional one to measure text formality based on the comments nature on such website:
\begin{itemize}
\item \textbf{Comments-content}: a set of features representing the bag-of-words (BOW) by each user, where each term represents a binary feature that is set to 1 if appeared in the user's comments and zero otherwise.
\item \textbf{Linguistic Inquiry and Word Count (LIWC)}: a set of features representing the percentage of terms used in user's comments that have a given LIWC category~\cite{LIWC:pennebaker2001linguistic}. This includes: emotions, questions, self reference, family reference \etc This set of features should show the nature of terms used by a given user, where it might be a good indication of the user's age and/or gender.
\item \textbf{Comments count}: This includes, \one~number of comments by the user; \two~percentage of unique comments, as we noticed many users repeats their exact comments on different posts; and \three~vocabulary variety, which is the number of unique terms used by the user divided by the total length of comments.
\item \textbf{Comments-formality}: One behaviour that young generations are commonly characterised by is the usage of slang, including shortcuts and abbreviations (\eg ``gr8'', ``thx''). We introduced a feature to measure the percentage of informal language in users' comments. To detect informal text, we used the Xerox part of speech (POS) tagger\footnote{\url{https://open.xerox.com/Services/fst-nlp-tools/}} to label POS tags of terms in comments. Terms with an undefined tag were considered informal language. We found this approach to be accurate in many cases. However, we noticed that some users use non-English terms. Thus, we applied the same POS tagger for three languages: English, French, and German. Those terms that obtained undefined tags with all three languages were considered informal. Finally, the feature value is the number of informal terms divided by the comments length.  
\end{itemize}

\subsubsection{Network features}
Another set of features were extracted to represent non-textual information. Most of these features are dependent on the network characteristics of PornHub, which contains more information on individuals than other general social websites. These features are:
\begin{itemize}
\item \textbf{Profile}: including a user's country, status (single / in a relationship), and interested in (men / women / both).
\item \textbf{Activity}: numbers of videos watched and posted (in log scale).
\item \textbf{Network}: Numbers of friends, subscribers, and subscribed to (in log scale). In addition, percentage of males and females in each of these lists.\footnote{Note that this feature should not be used in isolation, as it is partly a product of how others react to the profile's \emph{reported} demographics (a female account is likely to attract more attention even if it is fake).} 
\end{itemize}

Using the above features, we perform gender and age prediction to identify discrepancies, where users may be exhibiting anomalous behaviour for their stipulated demographics.

\subsection{Gender Prediction}
For gender prediction, we train a support vector machine (SVM) binary classifier~\cite{joachims2002learning} using the above features of the verified accounts. The created model allows us to then classify the remaining unverified accounts.

To measure the performance of the built classifier, we used 10-fold cross validation on the verified set. We only focused on the verified accounts with at least 10 comments in their profile to avoid training the classifier with samples of sparse content features. The number of verified accounts with 10+ comments are 1,231 out of which only 1,119 accounts had their gender identified. Numbers of males and females are 820 and 299 respectively. Precision, recall, and F-measure were calculated for each gender separately; then macro-F-measure and accuracy were calculated for the overall performance. 

Different combinations of the feature groups were tested to find the most effective features. Table~\ref{tab:GenderResults} reports the results for gender prediction on the verified accounts. As shown, network features are more effective than content features in predicting user's gender. This is driven by the differences seen in the popularity of the two genders (\S\ref{sec:char}). However, when combining content and network features, significantly better accuracy (92\%) is achieved. We note that the classifier is more effective in predicting males than females, which is expected since male training samples are almost three times than of female samples. This also means the impact of inacurracy will be inflated for female users pretending to be male. Nevertheless, we believe this still constitutes a good first step.

\subsection{Age Estimation}

The same set of features has been used for age estimation. We applied two regression techniques for age estimation; namely, SVM regression\footnote{\url{http://svmlight.joachims.org/}} and deep neural networks (DNNs).
We have examined different models for SVM, including linear and polynomial regressions. As for the neural network approach, we designed a multi-layered perceptron (MLP) trained to do regression over all the possible age values. The neural network had 3 hidden layers of rectified linear unit (ReLU) activated neurons each. This means the activation function for each neuron was of the form $f(x) = max(0,x)$

Our experimentation showed that SVM linear regression achieved the most effective results when compared to the other non-linear SVM regression models and DNNs. Actually, DNNs showed the poorest performance, which could be explained by over-fitting.
Unlike SVMs that are resilient to over-fitting because of their regularization techniques, DNNs are highly susceptible to over-fitting, especially in situations where feature dimensionality exceeds training dataset size. This is because of exponentially higher number of free parameters. We therefore utilise SVM.

Similar to what we performed in gender prediction, 10-fold cross-validation was applied on the verified accounts with 10+ comments to evaluate the effectiveness of the features and regression techniques. All the 1,231 verified accounts with 10+ comments were used, where the input for training the regression models was the extracted features set, and the label is the declared age by the verified account. We normalized ages over 60 to ``60+'' in order to reduce the sparsity for these ages. For measuring the performance, we followed the same method applied in literature~\cite{age:nguyen2011author,age:nguyen2013old,age:nguyen2014gender}, where Pearson's correlation and mean absolute error (MAE) between estimated age and actual age are measured.

\begin{table}[t]
\begin{center}									
{	
\begin{tabular}{l|cc}
Features & MAE & Correlation \\\hline
Content & \textbf{5.581} & \textbf{0.509} \\
Network & 6.119 & 0.234 \\
All & 5.783 & 0.440 \\\hline
\end{tabular}
\caption{Age estimation performance (SVM)} 
\label{tab:AgeResults}}	
\vspace{-0.5cm}
\end{center}
\end{table}

The best result we achieved with DNNs was a correlation of 0.28 and MAE of 6.8 years. This is much lower than the best results achieved with SVMs. Table~\ref{tab:AgeResults} shows the obtained MAE and correlation using different features set when applying SVM linear regression model for age estimation. Unlike gender detection, network features are not effective for age estimation. Actually, it degrades the performance of content features when added to it. Content features from users' comments are the most effective as shown in the table. The model can estimate the age of an account holder from his/her comments with an error of $\pm$5 years on average. The correlation between estimated ages and actual ones is 0.51. This result compares well with the state-of-the-art methods on other data sets~\cite{age:nguyen2011author}.\footnote{The best result for age estimation achieved in ~\cite{age:nguyen2011author} for medical blog users was a correlation of 0.535 and 6.537 MAE}

\section{Results and Discussion}
\label{sec:results}
Next, we utilise our classifiers to explore the unverified users who are classified as potential catfishes. We consider a catfish to be either: \one~a user who is classified as having a gender different to the one reported; or \two~a user with a predicted age that is more than 5.581 years\footnote{Our model achieves a MAE of 5.581. Thus, we cannot consider catfishes who only marginally change their age.} different than their claimed age.
Of course, this definition would also identify some verified accounts as catfishes in the case of classification errors. Consequently, we emphasise that our analysis highlights users who are \emph{potentially} catfishes --- it is not a ground truth. Hence, our classifiers should only be used for raising flags for applying additional vetting for users with suspicious behaviour.

For consistency, only unverified accounts with more than 10 comments were considered in our analysis, since our models were trained in the same way. The number of unverified accounts with 10+ comments included in our upcoming analysis is 11,182.

\subsection{Who catfishes?}

\begin{figure}
 	\centering
  	\includegraphics[width=0.98\columnwidth]{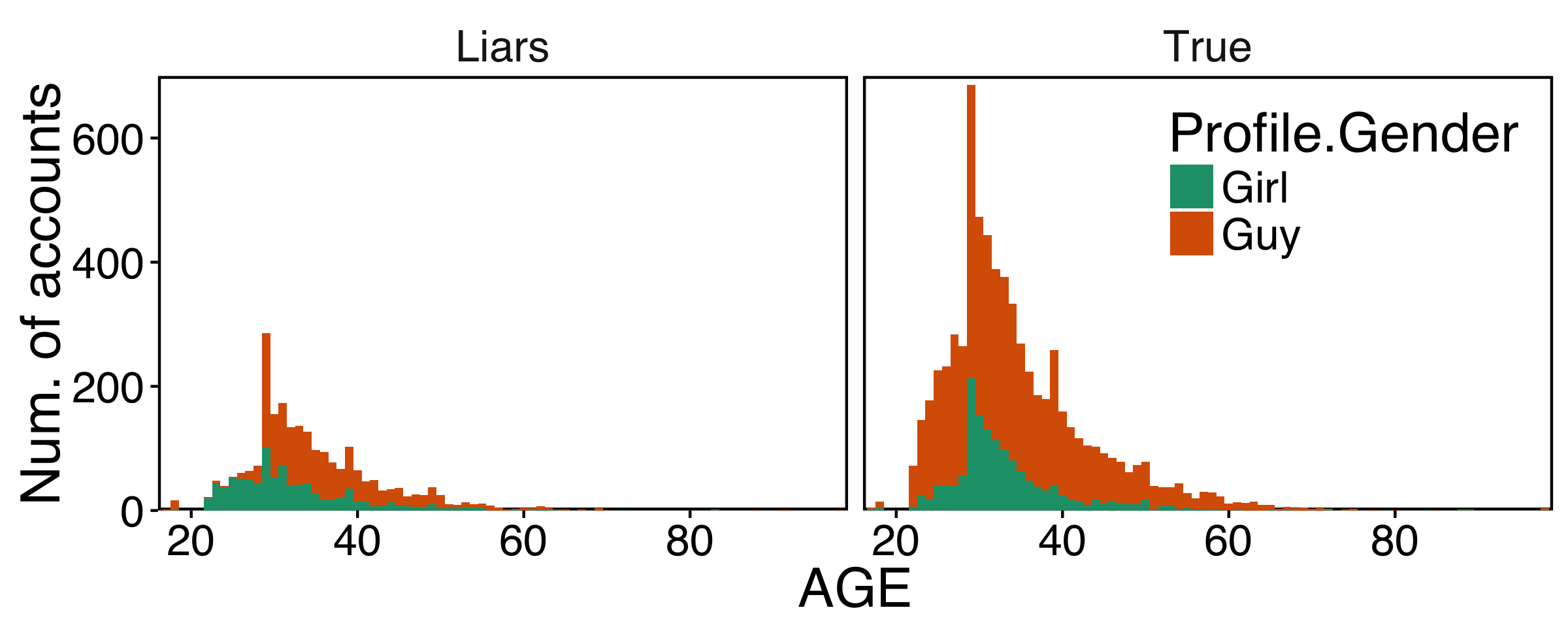}
    \caption{The age breakdown for users who lie about their gender (left) and those who don't (right).}
    \label{fig:age_for_liars}
    \vspace{-0.3cm}
\end{figure}

\begin{figure}
	\centering
 	\includegraphics[width=0.98\columnwidth]{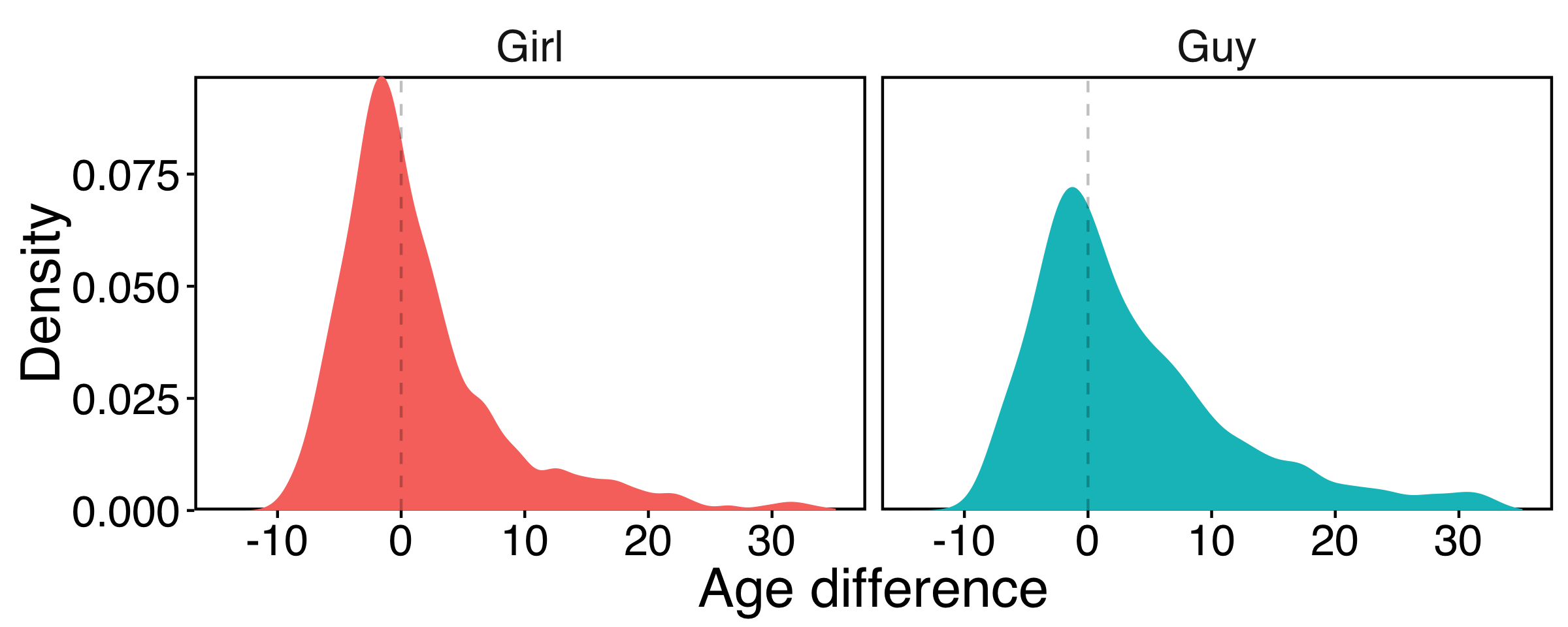}
    \caption{The probability density function of difference between estimated and reported user age, broken down for female and male accounts.}
        \label{fig:pred_age-diff_pdf}
        \vspace{-0.3cm}
\end{figure}

We start by inspecting the number of accounts that are classified as lying about their gender. Figure~\ref{fig:age_for_liars} shows the number of catfishes of either gender and in each age range (compared to the number of users classified as honest). Note that in Figure~\ref{fig:age_for_liars}, the gender refers to the \emph{reported} gender listed on the profile (not the one that was predicted by our classifier). A key observation emerges from Figure~\ref{fig:age_for_liars} --- a larger proportion of \emph{female} users pretend to be men. 
Just 35.1\% of male accounts are classified as catfishes compared to 60.7\% for women. 
Although this sounds high, it is roughly equivalent to prior questionnaire-based results (48\%~\cite{Cooper2000}).
It is also worth noting that a similar observation has been made in online gaming~\cite{martey2014strategic,lou2013}. A major reason for this was the perception that men are treated differently; we conjecture a similar motive could be at play here. We should also note that this may be driven by classifier errors caused by the disparity in male \vs female sample sizes; exploring this is a major line of our future work.

\begin{figure}
	\centering
 	\includegraphics[width=0.98\columnwidth]{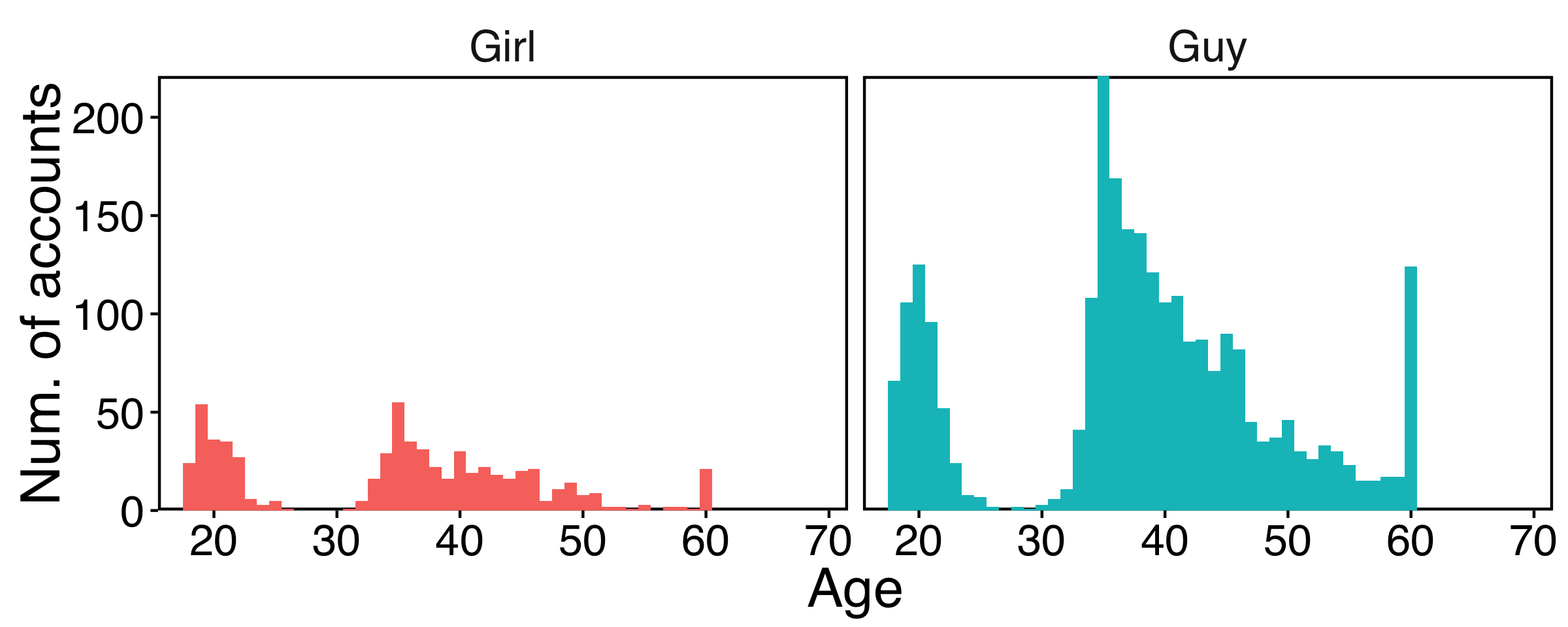}
    \caption{Reported age of accounts lying about their age, broken down by gender.}
        \label{fig:pred_age-age_gender}
\end{figure}

We also inspect accounts that are classified as lying about their age. Figure~\ref{fig:pred_age-diff_pdf} presents the difference between reported and predicted ages. It can be seen that catfish accounts reported as male (right) tend to select a marginally younger age, but also experiment with far older age ranges. In contrast, catfish accounts reported as female more consistently select younger ages.  The average age of a male catfish is 38.15, but just 34.87 for female catfish. 
This can clearly be seen in Figure~\ref{fig:pred_age-age_gender}, which presents a histogram of the reported ages for the accounts classified as catfishes: Male catfish profiles are more likely to be older than female catfish profiles.
This tendency is conducive with common social theory; studies have shown that women are generally interested in same-aged to somewhat older men, whereas men exhibit preferences towards younger women~\cite{Antfolk2015}. Hence, it seems likely that men pretending to be women would select young ages, whilst women pretending to be men would select from a wider range of ages.

\subsection{What are the benefits of catfishing?}
\begin{figure}
 	\centering
  	\includegraphics[width=0.98\columnwidth]{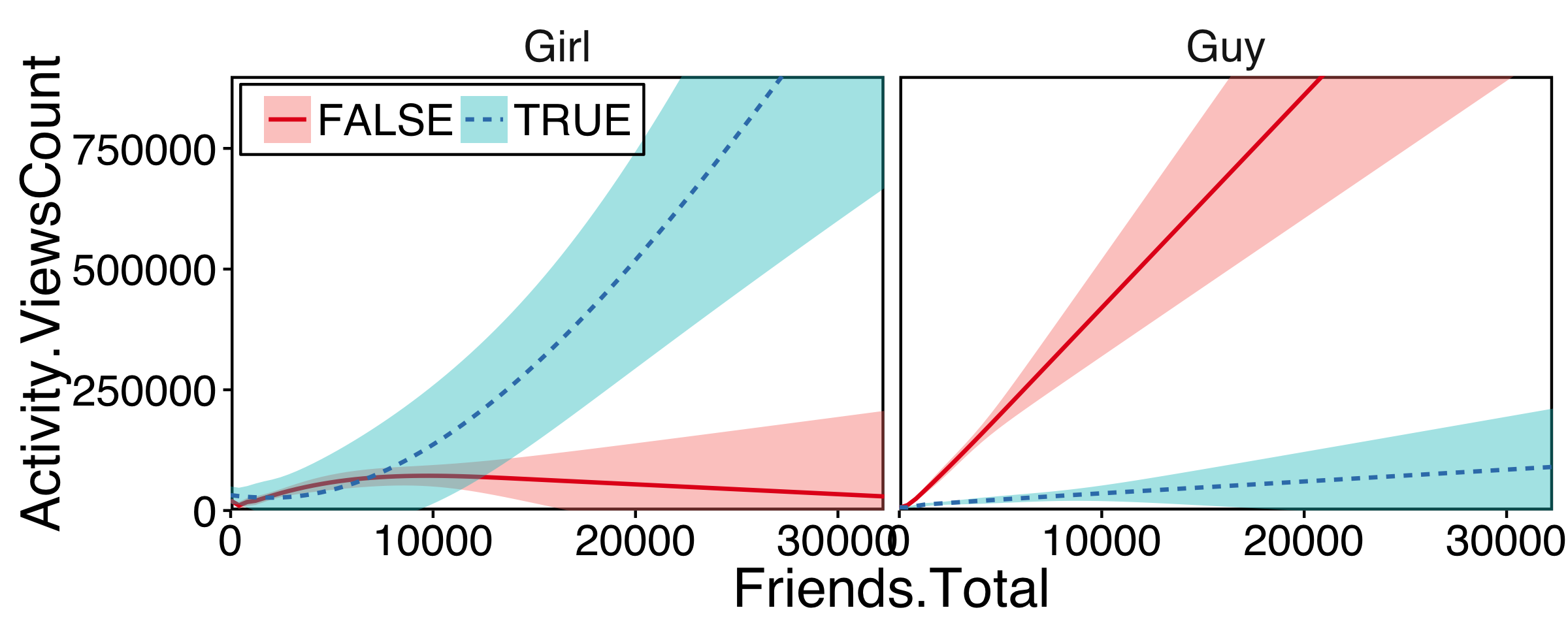}
    \caption{Number of Profile Views for profiles with false and true genders (smoothed with 95\% confidence interval). We use the classifier to separate accounts into male and female (rather than using the reported gender).}
    \label{fig:view_counts_for_liars}
\end{figure}

An obvious question is what drives users to catfish. Although there are many potential reasons, we have previously hypothesised that frustrated (unpopular) users may catfish in an attempt to gain more attention (\S\ref{sec:char}). This can be measured using the profile view counter listed on each profile, as well as the number of friends a profile garners. To inspect this, Figure~\ref{fig:view_counts_for_liars} plots the number of views \vs the number of friends. Profiles are separated into those classified as catfishes (false) and honest (true) users.
It can be seen that \emph{honest} (true) female accounts gain by far the most profile views and friend relationships. As the number of views increase, so does the number of friends. On average, female users have 32,150 views and 444 friends. 
More interesting, however, is comparing these against users classified as catfishes. 
Male catfishes, \ie male users with a female account (red line in the right plot), do substantially better and mirror the popularity of women (16590 views and 314 friends on average).
In contrast, the average view count for truthful male profiles is just 6098, while they gain only 195 friends.
Such observations shed light on a potential motivation of a male catfish. 
The situation is very different for female catfish, \ie women pretending to be men (red line in the left plot), who gain far fewer views and friends than their honest counterparts (8719 and 151 respectively). This is because, to other users, they will appear to be men --- clearly this raises further challenges regarding the classification process.


\begin{figure}
 	\centering
  	\includegraphics[width=0.98\columnwidth]{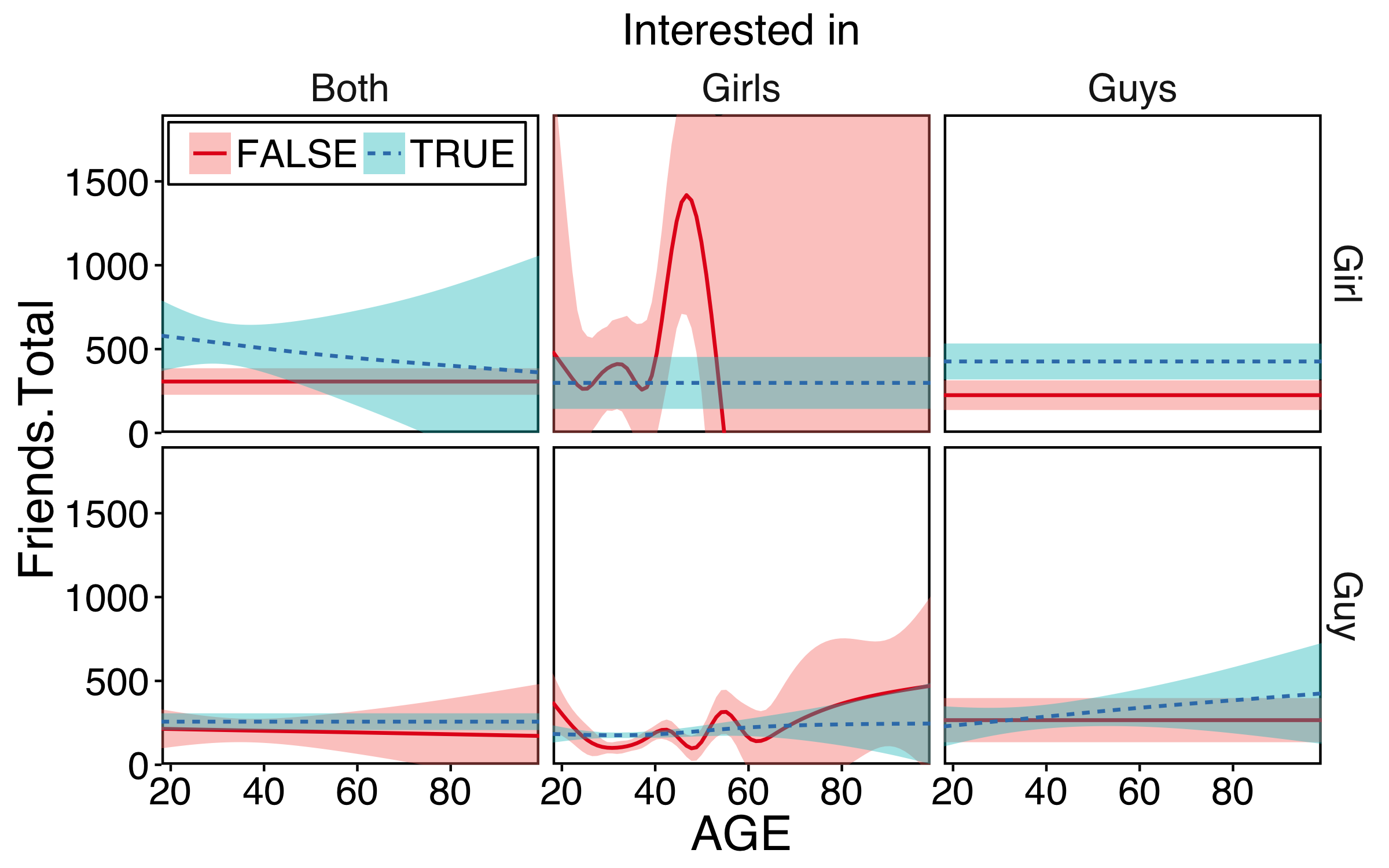}
    \caption{The number of friends per account as a proxy gain for lying about gender, broken down by gender and interest.}
    \label{fig:pred_gender-frnds-age_interest}
\end{figure}

\subsection{What do catfishes want?}
We can also decompose the profiles into accounts interested in each option (men, women, both). This allows us to see what profiles classified as catfishes are typically seeking. Figure~\ref{fig:pred_gender-frnds-age_interest} plots the number of friends each account accumulated based on the user's age. It also breaks down users based on what they stipulate they are interested in.
It can be seen that very few male profiles register an interest in both genders (8.0\%), whilst this is significantly higher for female accounts (60.5\%). We conjecture this might be because females choose to catfish so that they can experiment with other women whilst also interacting with men, whilst men choose to catfish primarily so they can overcome the difficulty of befriending women.
Hence, male catfishes tend to lie about being homosexual women; these are actually the users who gain the most (in terms of increased friendship sizes). 
This is especially the case for men aged 30-55 years (red line in top middle plot), who tend to do quite poorly when being truthful.

\section{Conclusion}
\label{sec:conc}

In this preliminary study, we have explored the behaviour and characteristics of users in a major Porn 2.0 social network. Our focus has been on understanding the roles of age and gender, exploiting state-of-the-art algorithms to try and detect users who potentially lie about these features: so called ``catfishes''. To achieve our goal, we built classifiers for gender and age estimation. Again, the predictions remain unverified by a ground truth and we therefore temper our findings with this consideration. The classifier suggested a number of trends, \eg that women actually have a greater propensity to lie than men. The classifier also suggested that the two genders tend to select different ages when catfishing. Whereas men pretending to be women usually select younger ages, women pretending to be men select from a wider range.

This work is just the first step towards understanding the nature of online catfishing. Here, we have only explored deception related to age and gender. Our future work will expand this to understand how some users lie about other attributes such as location and interests. We also plan to use machine learning to better model the similarities between some male and female accounts (\eg based on browsing patterns~\cite{Elkhatib2014browsing}). The most important priority is confirming our findings via ground truth information (\eg from surveys and questionnaires). This is because it is currently impossible to gain definitive data beyond the verified profiles. Our method is thus underpinned by the assumption that there is limited catfishing related to gender and age in verified profiles. We acknowledge that this assumption likely contains flaws, thereby introducing noise to our findings. That said, we argue that this is the first significant dataset to allow such exploration. The classifier also assumes that the truthful underlying attributes of unverified profiles are similar to that of verified profiles. This too needs much more testing before the results can be considered conclusive. Finally, we wish to understand \emph{why} users choose to lie, and how individual users strategise this deception. Again, we plan to execute this via surveys that can garner deeper insight into such reasoning.

\bibliographystyle{IEEEtran}
\balance{\bibliography{ref}}

\begin{thebibliography}{10}
\providecommand{\url}[1]{#1}
\csname url@samestyle\endcsname
\providecommand{\newblock}{\relax}
\providecommand{\bibinfo}[2]{#2}
\providecommand{\BIBentrySTDinterwordspacing}{\spaceskip=0pt\relax}
\providecommand{\BIBentryALTinterwordstretchfactor}{4}
\providecommand{\BIBentryALTinterwordspacing}{\spaceskip=\fontdimen2\font plus
\BIBentryALTinterwordstretchfactor\fontdimen3\font minus
  \fontdimen4\font\relax}
\providecommand{\BIBforeignlanguage}[2]{{%
\expandafter\ifx\csname l@#1\endcsname\relax
\typeout{** WARNING: IEEEtran.bst: No hyphenation pattern has been}%
\typeout{** loaded for the language `#1'. Using the pattern for}%
\typeout{** the default language instead.}%
\else
\language=\csname l@#1\endcsname
\fi
#2}}
\providecommand{\BIBdecl}{\relax}
\BIBdecl

\bibitem{Cooper2000}
A.~Cooper, D.~L. Delmonico, and R.~Burg, ``Cybersex users, abusers, and
  compulsives: New findings and implications,'' \emph{Sexual Addiction and
  Compulsivity}, vol.~7, no. 1-2, pp. 5--29, 2000.

\bibitem{tyson2015people}
G.~Tyson, Y.~Elkhatib, N.~Sastry, and S.~Uhlig, ``Are people really social on
  porn 2.0?'' in \emph{ICWSM}, 2015.

\bibitem{age:nguyen2011author}
D.~Nguyen, N.~A. Smith, and C.~P. Ros{\'e}, ``Author age prediction from text
  using linear regression,'' in \emph{Proceedings of the 5th ACL-HLT Workshop
  on Language Technology for Cultural Heritage, Social Sciences, and
  Humanities}.\hskip 1em plus 0.5em minus 0.4em\relax Association for
  Computational Linguistics, 2011, pp. 115--123.

\bibitem{age:nguyen2014gender}
D.-P. Nguyen, R.~Trieschnigg, A.~Do{\u{g}}ru{\"o}z, R.~Gravel, M.~Theune,
  T.~Meder, and F.~de~Jong, ``Why gender and age prediction from tweets is
  hard: Lessons from a crowdsourcing experiment.''\hskip 1em plus 0.5em minus
  0.4em\relax Association for Computational Linguistics, 2014.

\bibitem{carroll2008generation}
J.~S. Carroll, L.~M. Padilla-Walker, L.~J. Nelson, C.~D. Olson, C.~M. Barry,
  and S.~D. Madsen, ``Generation {XXX} pornography acceptance and use among
  emerging adults,'' \emph{Journal of adolescent research}, vol.~23, no.~1,
  2008.

\bibitem{daneback2012outcomes}
K.~Daneback, A.~Sevcikova, S.-A. M{\aa}nsson, and M.~W. Ross, ``Outcomes of
  using the internet for sexual purposes: fulfilment of sexual desires,''
  \emph{Sexual Health}, vol.~10, no.~1, 2012.

\bibitem{Tyson:2016}
G.~Tyson, Y.~Elkhatib, N.~Sastry, and S.~Uhlig, ``Measurements and analysis of
  a major adult video portal,'' \emph{ACM Trans. Multimedia Comput. Commun.
  Appl.}, vol.~12, no.~2, 2016.

\bibitem{Fiadino13}
P.~Fiadino, A.~B{\"a}r, and P.~Casas, ``{HTTPTag}: A flexible on-line {HTTP}
  classification system for operational {3G} networks,'' in \emph{INFOCOM},
  2013.

\bibitem{age:buzzell2005demographic}
T.~Buzzell, ``Demographic characteristics of persons using pornography in three
  technological contexts,'' \emph{Sexuality \& Culture}, vol.~9, no.~1, pp.
  28--48, 2005.

\bibitem{age:marquardt2014age}
J.~Marquardt, G.~Farnadi, G.~Vasudevan, M.-F. Moens, S.~Davalos, A.~Teredesai,
  and M.~De~Cock, ``Age and gender identification in social media,'' in
  \emph{Proc. of CLEF 2014 Evaluation Labs}, 2014.

\bibitem{age:rangel2013use}
F.~Rangel and P.~Rosso, ``Use of language and author profiling: Identification
  of gender and age,'' \emph{Natural Language Processing and Cognitive
  Science}, vol. 177, 2013.

\bibitem{age:nguyen2013old}
D.-P. Nguyen, R.~Gravel, R.~Trieschnigg, and T.~Meder, ````{H}ow old do you
  think {I} am?'' {A} study of language and age in {Twitter},'' 2013.

\bibitem{age:al2012homophily}
F.~Al~Zamal, W.~Liu, and D.~Ruths, ``Homophily and latent attribute inference:
  Inferring latent attributes of {Twitter} users from neighbors,''
  \emph{ICWSM}, vol. 270, 2012.

\bibitem{UKporn}
\BIBentryALTinterwordspacing
J.~Killock, ``A database of the uk's porn habits. what could possibly go
  wrong?'' \emph{Open Right Group}, Oct. 2016. [Online]. Available:
  \url{https://www.openrightsgroup.org/blog/2016/a-database-of-the-uks-porn-habits-what-could-possibly-go-wrong}
\BIBentrySTDinterwordspacing

\bibitem{attwood2010porn}
F.~Attwood, \emph{Porn.com: Making sense of online pornography}.\hskip 1em plus
  0.5em minus 0.4em\relax Peter Lang, 2010, vol.~48.

\bibitem{schuhmacher9exploring}
M.~Schuhmacher, C.~Zirn, and J.~V{\"o}lker, ``Exploring {YouPorn} categories,
  tags, and nicknames for pleasant recommendations,'' in \emph{Proc. Workshop
  on Search and Exploration of X-Rated Information}, 2013.

\bibitem{jacobs2009}
K.~Jacobs, ``Is there life on {Adult} {FriendFinder}? {S}ex and logic with the
  happy dictator,'' in \emph{Proceedings of the Digital Arts and Culture
  Conference}, 2009.

\bibitem{tyson2016first}
G.~Tyson, V.~C. Perta, H.~Haddadi, and M.~C. Seto, ``A first look at user
  activity on tinder,'' in \emph{IEEE/ACM International Conference on Advances
  in Social Networks Analysis and Mining (ASONAM)}, 2016.

\bibitem{trestian2013}
I.~Trestian, C.~Xiao, and A.~Kuzmanovic, ``A glance at an overlooked part of
  the world wide web,'' in \emph{Proc. WWW Conference}, 2013.

\bibitem{Tyson13}
G.~Tyson, Y.~Elkhatib, N.~Sastry, and S.~Uhlig, ``Demystifying porn 2.0: a look
  into a major adult video streaming website,'' in \emph{Proc.\ IMC}, 2013.

\bibitem{Hurley13}
R.~Hurley, S.~Prusty, H.~Soroush, R.~J. Walls, J.~Albrecht, E.~Cecchet, B.~N.
  Levine, M.~Liberatore, B.~Lynn, and J.~Wolak, ``{Measurement and Analysis of
  Child Pornography Trafficking on {P2P} Networks},'' in \emph{Proc. WWW
  Conference}, 2013.

\bibitem{bruckman1996gender}
A.~Bruckman, ``Gender swapping on the internet,'' \emph{High noon on the
  electronic frontier: Conceptual issues in cyberspace}, pp. 317--326, 1996.

\bibitem{martey2014strategic}
R.~M. Martey, J.~Stromer-Galley, J.~Banks, J.~Wu, and M.~Consalvo, ``The
  strategic female: gender-switching and player behavior in online games,''
  \emph{Information, Communication \& Society}, vol.~17, no.~3, 2014.

\bibitem{lou2013}
J.-K. Lou, K.~Park, M.~Cha, J.~Park, C.-L. Lei, and K.-T. Chen, ``Gender
  swapping and user behaviors in online social games,'' in \emph{Proc. WWW
  Conference}, 2013.

\bibitem{hussain2008gender}
Z.~Hussain and M.~D. Griffiths, ``Gender swapping and socializing in
  cyberspace: An exploratory study,'' \emph{CyberPsychology \& Behavior},
  vol.~11, no.~1, pp. 47--53, 2008.

\bibitem{almishari2016trilateral}
M.~Almishari, E.~Oguz, and G.~Tsudik, ``Trilateral large-scale osn account
  linkability study,'' in \emph{AAAI Fall Symposium Series}, 2016.

\bibitem{6459494}
A.~Rashid, A.~Baron, P.~Rayson, C.~May-Chahal, P.~Greenwood, and J.~Walkerdine,
  ``Who am i? analyzing digital personas in cybercrime investigations,''
  \emph{Computer}, vol.~46, no.~4, pp. 54--61, April 2013.

\bibitem{rowe2016mining}
M.~Rowe and H.~Saif, ``Mining pro-{ISIS} radicalisation signals from social
  media users,'' in \emph{ICWSM}, 2016, pp. 329--338.

\bibitem{ferree2003women}
M.~Ferree, ``Women and the web: Cybersex activity and implications,''
  \emph{Sexual and Relationship Therapy}, vol.~18, no.~3, pp. 385--393, 2003.

\bibitem{LIWC:pennebaker2001linguistic}
J.~W. Pennebaker, M.~E. Francis, and R.~J. Booth, ``Linguistic inquiry and word
  count ({LIWC}): A computerized text analysis program,'' \emph{Mahwah (NJ)},
  vol.~7, 2001.

\bibitem{joachims2002learning}
T.~Joachims, \emph{Learning to classify text using support vector machines:
  Methods, theory and algorithms}.\hskip 1em plus 0.5em minus 0.4em\relax
  Kluwer Academic Publishers, 2002.

\bibitem{Antfolk2015}
J.~Antfolk, B.~Salo, K.~Alanko, E.~Bergen, J.~Corander, N.~K. Sandnabba, and
  P.~Santtila, ``Women's and men's sexual preferences and activities with
  respect to the partner's age: evidence for female choice,'' \emph{Evolution
  and Human Behavior}, vol.~36, no.~1, pp. 73 -- 79, 2015.

\bibitem{Elkhatib2014browsing}
Y.~Elkhatib, R.~Killick, M.~Mu, and N.~Race, ``{Just browsing? Understanding
  user journeys in online TV},'' in \emph{International Conference on
  Multimedia}.\hskip 1em plus 0.5em minus 0.4em\relax ACM, 2014, pp. 965--968.

\end{thebibliography}

\end{document}